# Exploring the limits of pre-trained embeddings in machine-guided protein design: a case study on predicting AAV vector viability

Ana F. Rodrigues[1,*,✉], Lucas Ferraz[1,*], Laura Balbi[1], Pedro Giesteira Cotovio[1], Catia Pesquita[1]

1 - LASIGE, Faculdade de Ciências da Universidade de Lisboa, Lisboa, Portugal

* – Equal contribution | ✉ – Correspondence should be addressed to A.F.R. (afdrodrigues@fc.ul.pt)

## Abstract

Effective representations of protein sequences are widely recognized as a cornerstone of machine learning-based protein design. Yet, protein bioengineering poses unique challenges for sequence representation, as experimental datasets typically feature few mutations, which are either sparsely distributed across the entire sequence or densely concentrated within localized regions. This limits the ability of sequence-level representations to extract functionally meaningful signals. In addition, comprehensive comparative studies remain scarce, despite their crucial role in clarifying which representations best encode relevant information and ultimately support superior predictive performance. In this study, we systematically evaluate multiple ProtBERT and ESM2 embedding variants as sequence representations, using the adeno-associated virus capsid as a case study and prototypical example of bioengineering, where functional optimization is targeted through highly localized sequence variation within an otherwise large protein. Our results reveal that, prior to fine-tuning, amino acid–level embeddings outperform sequence-level representations in supervised predictive tasks, whereas the latter tend to be more effective in unsupervised settings. However, optimal performance is only achieved when embeddings are fine-tuned with task-specific labels, with sequence-level representations providing the best performances. Moreover, our findings indicate that the extent of sequence variation required to produce notable shifts in sequence representations exceeds what is typically explored in bioengineering studies, showing the need for fine-tuning in datasets characterized by sparse or highly localized mutations.

# 1. Introduction

Machine learning (ML)-based protein design has become a powerful strategy in modern protein engineering [1]. A critical aspect of this approach is selecting an appropriate format to represent the protein sequence as input for the ML model. The optimal representation depends on factors such as the specific task to be learnt and dataset characteristics (for comprehensive reviews on the topic, see, for example, Yue *et al.* (2023) [2] or Harding-Larsen *et al.* (2024) [3]). Since different representations capture distinct types of information and give rise to different data structures and properties, this decision can significantly impact computational complexity, model accuracy, and generalizability.

A relevant use case where optimal representation formats are critical is the ML-based protein bioengineering of viral vectors for gene therapy. The datasets used in this field tend to be either deep mutational scanning studies, which target the entire protein sequence but typically introduce only mutation per variant (e.g. [4]), or high-intensity mutational studies, which explore a dense set of variations but restricted to small regions of the protein. For example, in recent ML-based bioengineering studies of adeno-associated virus (AAV) vectors, researchers have typically concentrated on a 20–50 amino acid fragment region of the capsid, which exceeds 700 residues [5–7]. This challenge extends beyond viral vectors and reflects a general issue in protein bioengineering studies, including in areas such as therapeutic enzyme optimization [8] and peptides or antibody design for cancer [9,10], where sequence variants rarely contain dense mutations spanning the entire length. This has important implications for representation formats since many of them, including those considered state-of-the-art, are designed to capture broad sequence-level information at the scale of the entire protein. When functional optimization depends instead on small or highly localized sequence changes, these representations may dilute the signal of interest within a much larger context. Therefore, it is crucial to evaluate how the number and localization of sequence changes affect the various representation formats, how they perform under these constraints, and to identify strategies to adapt them for data regimes characteristic of this field.

Protein representations can be broadly categorized into three main groups, based on the type of information they capture: primary amino acid sequence, three-dimensional structure, and molecular dynamics or activity [3]. Multimodal approaches that integrate various types of information are also gaining traction [11]. Representations based on primary amino acid sequences are often the first choice, as sequence data is more readily available and offers greater reliability. Traditionally, sequences are transformed into numerical formats based on predefined rules, such as positional encoding schemes (e.g., one-hot encoding (OHE) of amino acids) or descriptors that summarize physicochemical properties of amino acid residues (e.g., hydrophobicity, charge, or molecular weight) [12]. These formats are relatively simple to generate and have long served as a workhorse for classical ML approaches. However, their reliance on hand-crafted features requires the user to decide what information is most relevant, which can limit flexibility and bias the model with arbitrary assumptions. More recently, representations learnt directly from raw amino acid sequences, called embeddings, have gained relevance [11].

An embedding is a continuous vector that represents an entity (e.g., a protein, word, image). Embeddings can be generated using several methods, but modern approaches typically rely on deep learning models that learn representations from large datasets by mapping raw inputs into high-dimensional feature spaces. This enables features to be learned automatically and allows the representation to encode rich, high-level information beyond raw data and without explicit user intervention [13]. Such representations are generally referred to as pre-trained embeddings, as they are learned prior to their application in specific downstream tasks. Because embeddings represent entities in the same n-dimensional space, they support similarity calculation between entities [13]. Another key advantage of embeddings is their adaptability through fine-tuning, a transfer learning strategy in which a pre-trained model is further trained on a smaller, task-specific dataset [14]. This process tailors the embeddings to the downstream task (e.g., protein function classification) by refining them to capture task-relevant patterns, enhancing their specificity, discriminative power, and overall predictive accuracy [15].

Among embeddings, those generated with transformer-based models [16] have the added value of incorporating attention mechanisms [17]. Attention enables the model to dynamically focus on different regions of the input sequence when generating the embedding. For proteins, this allows the model to capture complex, short- and long-range dependencies within the sequence that are critical for understanding structural and functional relationships, resulting in context-sensitive representations that reflect interactions between distant amino acid residues. Within transformer-based protein language models (pLMs), TAPE [18] was a pioneer. However, it has since been outperformed by newer architectures trained on larger sequence datasets, with greater model capacities and enhanced pretraining objectives, leading to superior performance across diverse downstream tasks (e.g., Capet *et al.* (2022) [19]). Currently, ESM and ProtBERT stand out as the most popular with applications in predicting protein function [20,21], structure modeling [22,23], or protein–protein interactions [24,25]. ESM-based models [26–28] are widely recognized for their state-of-the-art accuracy, particularly in structure-related tasks, although the most advanced versions tend to be computationally intensive. ProtBERT [29] offers a more lightweight and accessible option, which is especially appealing for settings with limited computational resources.

ProtBERT is based on the Google's Bidirectional Encoder Representations from Transformers (BERT) model [30], initially developed for natural language processing (NLP) using the encoder component of the Transformer architecture introduced by Vaswani *et al.* (2017) [16]. ProtBERT adapts this architecture to model protein sequences by treating proteins as sentences and amino acids as tokens. It is pre-trained on large protein sequence databases using masked language modeling, a self-supervised approach in which a fraction of amino acids in each sequence are randomly masked and the model learns to predict them from their surrounding context. Embeddings generated by ProtBERT have been demonstrated to capture rich and biologically meaningful information encoding key aspects of protein biology, including [31]: i) the ability to model protein folding by capturing long-range dependencies between amino acids that are distant in sequence but close in 3D space, ii) the identification of functionally important regions such as active or binding sites, and iii) the representation of complex biophysical and biochemical properties such as solvent accessibility, secondary structure, hydrophobicity, and evolutionary conservation. ESM2[27], the ESM-based model most directly comparable to ProtBERT, is a large-scale protein language model specifically designed for protein sequences. Like ProtBERT, it is trained using masked language modeling, but ESM2 employs deeper and larger Transformer architectures trained on hundreds of millions of sequences, thus, with the potential to yield more expressive and biologically informative embeddings.

The richness and completeness of information of pre-trained pLM embeddings surpasses that provided by traditional positional-encoding-based representation formats, placing these representation formats in a more favorable position to capture sequence-to-function relationships, and making them a prime choice for protein design and bioengineering use cases. However, this presumed advantage is not universally established. Some studies report improved performance when using pre-trained embeddings [32] over more traditional representations, while others find the benefits less clear [33]. These conflicting results likely depend on task type, model architecture, and dataset characteristics. The latter includes the challenge of learning from sequence variants with minor changes. Therefore, expanding comparative studies is essential to clarify when and why particular representation formats succeed. Additionally, pLMs can generate different variants of embeddings. While some studies have begun to explore their potential, the contexts in which each variant is most effective are not yet fully understood and would benefit from further comparative investigation.

Here, we systematically evaluate several embedding variants from ProtBERT and ESM2 for both unsupervised and supervised learning. We use AAV vector capsid bioengineering as a case study due to its significance in gene therapy [34], ongoing efforts in capsid bioengineering [35], and the availability of an extensive, experimentally validated dataset with relevant functional labels where mutations concentrate in a very small region of the capsid [5]. This setting provides an ideal testbed for examining the limits of pre-trained embeddings in capturing highly localized sequence changes. We further explore end-to-end fine-tuning as a strategy to overcome these limitations and enable protein bioengineering to benefit from the value of embedding-based representations.

## 2. Results

This study investigates how pre-trained embeddings perform as representations of protein sequences in ML tasks relevant to protein bioengineering, particularly when datasets contain only small and localized sequence changes, a common scenario in this field. We examine different embedding variants that can be generated directly by ProtBERT and ESM2 (Figure 1A): i) the global sequence embedding, which captures sequence-level information via the CLS (classification) token output from the final transformer layer; and ii) the amino acid embedding, which corresponds to the mean-pooled representation of all amino acids tokens in the sequence, excluding the special tokens CLS and SEP (separation); in the case of ProtBERT, it is also possible to obtain a third variant, the projected sequence embedding, which applies a dense fully connected layer followed by a tanh activation to the global sequence embedding. These formats were assessed in both supervised and unsupervised learning, and the results were analyzed to gain insight into the biological features each embedding type tends to capture (Figure 1B). OHE was included as a traditional baseline for benchmarking. Finally, supervised fine-tuning was explored as a strategy to adapt general-purpose embeddings to task-specific contexts.

Our case study focuses on the capsid protein of AAV2, the central structural component of the viral particle and a major target of machine-guided viral vector bioengineering efforts[4–7,36–39]. AAVs provide a representative and biotechnology relevant system in which major functional changes can be driven by small, localized sequence modifications. Although the full protein is 735 amino acids long, the dataset sequences used here contain changes exclusively within amino acids 561–588, comprising single and multiple mutations, including substitutions, deletions, and insertions, either alone or in combination. This region is frequently targeted in AAV bioengineering because it plays a key role in determining the vector's tropism, immune response, and overall stability [40], making it a prime candidate for modifications to optimize therapeutic potential. The dataset itself (viability set from Bryant *et al.* (2021) [5]) contains over 293,000 unique sequences experimentally tested for their ability to produce viable AAV particles (Supplementary Table 1), making it one of the most extensive resources available with experimental validation on AAV2 viability. In addition, the sequences were generated using a diverse range of machine-guided design strategies, including both ML-based and non-ML-based. This diversity provides multiple functional perspectives, enabling a richer, multi-faceted evaluation of different representation formats.

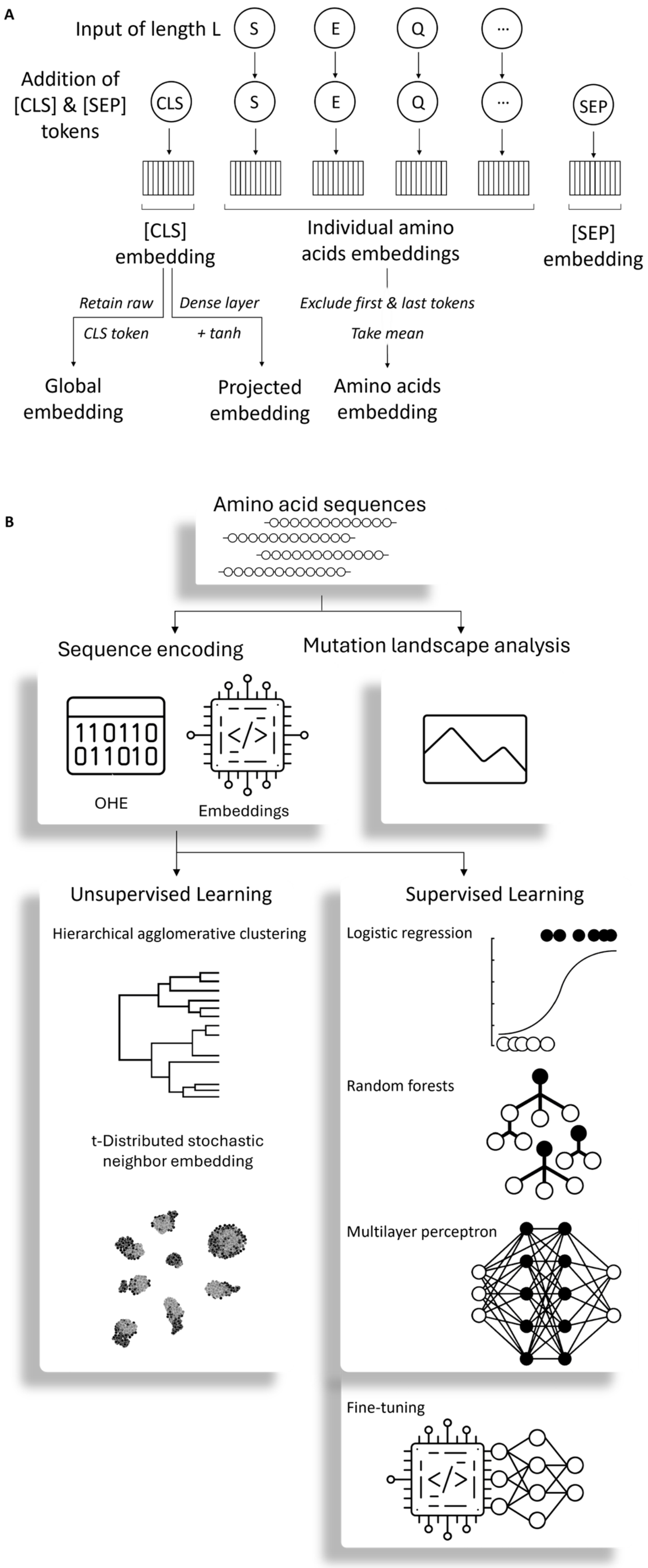

**Figure 1.** (**A**) Common embedding types generated by protein language models. A protein sequence, such as 'SEQ', is split into individual amino acid tokens. Two special tokens, CLS (classification) and SEP (separation) or EOS (end of sequence), are also added. Tokens are fed into the model, which produces *k*-dimensional embeddings typically using the last hidden layer. The two most common embedding variants often used are: i) the global sequence embedding, which captures the overall sequence meaning using the raw CLS token, and ii) the amino acids embedding, which is the average embedding of all amino acids. In the case of ProtBERT, it is also possible to extract the projected embedding, which is the raw CLS token embedding after going through a dense layer and tanh activation. (**B**) Overview of the main methodological approaches used in this work. This figure was created using icons from The Noun Project, used under the CC BY 3.0 license: *Binary Code* by Ahmad Arzaha, *Embedding* by Vectors Point, *Hierarchical Clustering Analysis*, *Binary Logistic Regression*, and *Deep Learning* by Product Pencil, *unnamed (Landscape)* by dilayorganci, *Artificial Neural Network* and *unnamed (Random Forests)* by sachin modgekar, and *Neural Network* by karyative.

### 2.1. Unsupervised learning

To assess the effectiveness of the different ProtBERT and ESM2 embeddings, we first use them in classical unsupervised learning techniques with strong visualization capabilities, namely hierarchical agglomerative clustering (HAC, Figure 2) and t-distributed stochastic neighbor embedding (t-SNE, Figure 3). These are commonly used in protein design and bioengineering studies, both in early stages as exploratory data analysis tools and later for interpreting model outputs and guiding design decisions. t-SNE [41] is particularly effective for interpreting high-dimensional spaces, such as those of embeddings, which, while mathematically rich, are often opaque to direct human understanding. It bridges abstract vector representations and human intuition, enabling researchers to visually assess whether embeddings cluster according to known biological or functional properties, and to detect meaningful subgroups. HAC [42] complements t-SNE by providing a hierarchical view of similarity across sequences, enabling the identification of nested clusters at multiple levels of granularity. This is particularly helpful in uncovering both broad and fine-scale groupings, such as functionally distinct variants within a single group. These methods were used to evaluate how well the representations could distinguish between known ground truths in the dataset, specifically, sequence viability (viable or non-viable) and the design strategy used to generate the sequences (ML-designed or non-ML-designed). In the case of t-SNE, the ML-designed group was further resolved according to the ML model type used for sequence design.

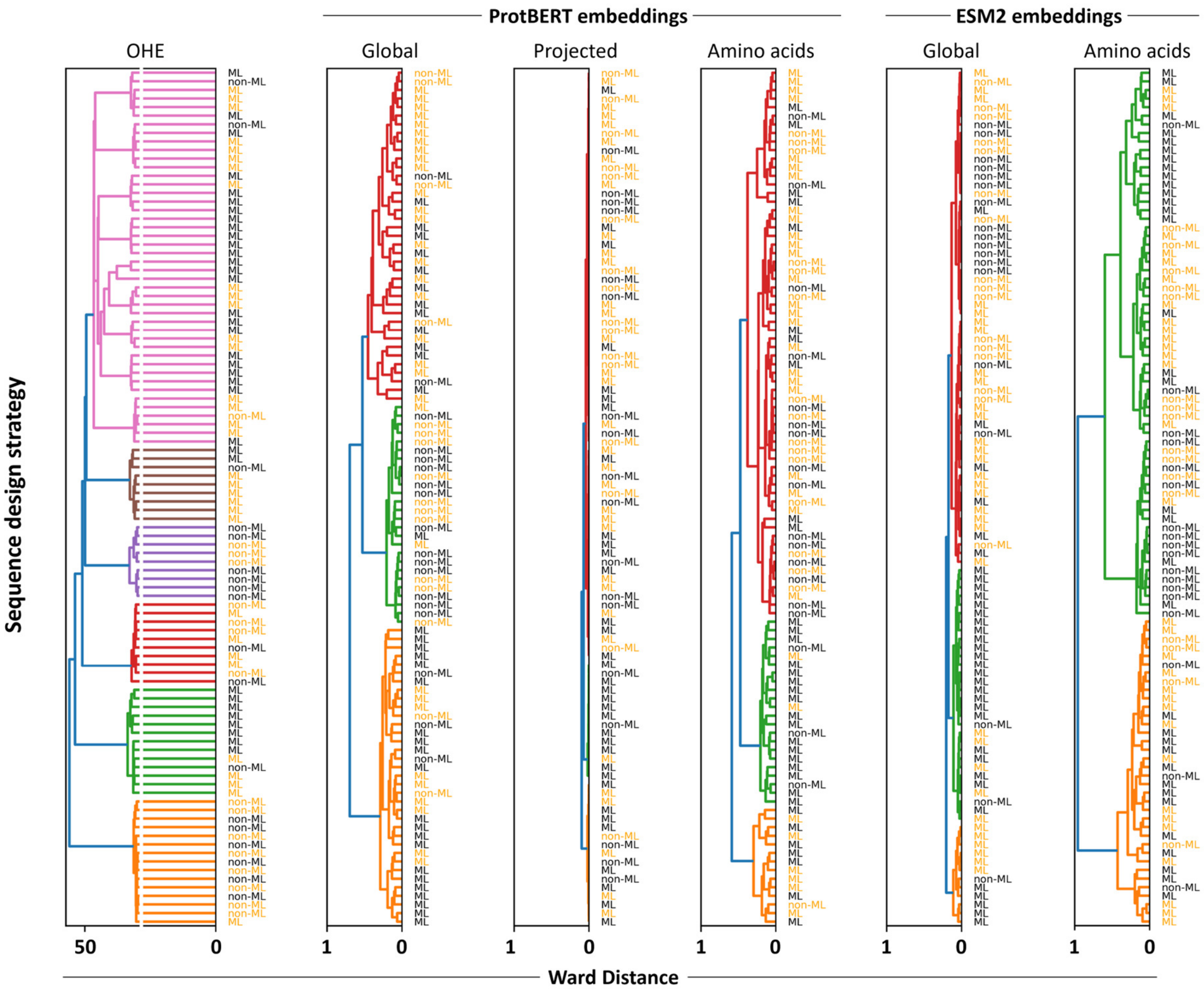

**Figure 2.** Hierarchical clustering of the different representation formats evaluated in this study, colored by viability (orange = viable, black = non-viable), and annotated by design strategy (ML- or non-ML-designed sequences). Note the scale differences between OHE and embedding-based representations.

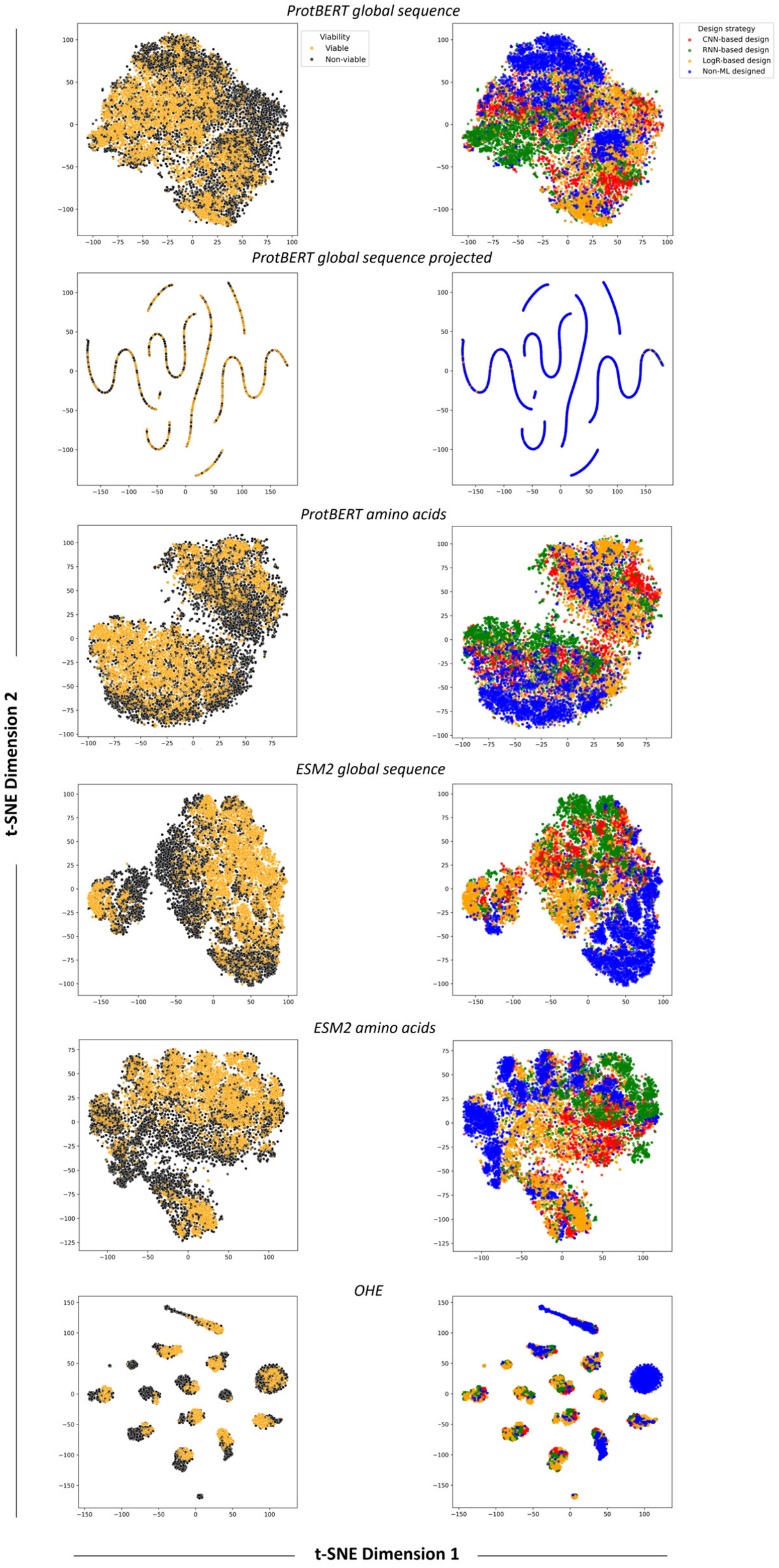

**Figure 3.** t-SNE plots of the different representation formats, colored by viability (left panels) or by design strategy (right panels).

In the HAC analysis, the OHE representation yielded much larger distances between sequences, with Ward linkage values exceeding 50 and six clusters appearing near the root of the dendrogram. By contrast, embeddings produced far smaller inter-sequence distances (below 1) and only two or three clusters close to the root. Clusters from OHE were also the most homogeneous in design strategy, although this is likely more due to the higher number of clusters rather than better discriminative power. All representation types demonstrated a reasonable ability to group sequences by design strategy, with projected embeddings consistently showing the weakest performance; however, none achieved complete separation of all classes. Similar patterns emerged in the t-SNE plots: no representation format displayed clear clustering based on viability, although most showed some aggregation by design strategy, except for projected embeddings, which remained the poorest performers. Notably, the t-SNE visualization of the OHE data revealed well-defined clusters, which were found to be highly homogeneous in terms of sequence length (Supplementary Figure 1). This was maintained after padding and SVD transformation, where the features in representation format no longer correlate with sequence size, and was also evident in HAC (Supplementary Figure 1). In contrast, pLM embeddings showed no clear clustering by sequence length. This suggests that OHE captures sequence size as a key feature, supporting that the most significant driver of clustering when using this representation format is neither the viability of the sequences nor the design strategy. This observation prompted us to investigate whether sequence size alone could predict ground-truth labels (Supplementary Figure 2). The analysis showed that, except for very large sequences (more than 747 amino acids), sequence size was a poor predictor of viability. However, sequences longer than 738 amino acids were predominantly associated with ML-based design strategies, indicating that in this case sequence length may have served as a stronger proxy for design origin, potentially making this label easier to distinguish.

### 2.2. Supervised learning

After evaluating the performance of different representations in unsupervised learning, the focus shifted to supervised learning, specifically the classification of sequence viability using three different ML model types: random forests, logistic regression, and a multilayer perceptron (MLP). In this context, using OHE in its binary format was not feasible, as the resulting feature space exceeded 15,000 dimensions, which limits efficiency and makes the training process computationally impractical. To address this, we applied truncated singular value decomposition (SVD) to generate a more compact representation of OHE vectors. The number of retained singular vectors was set to match the dimensionality of the embeddings (1024 for ProtBERT and 1280 for ESM2) to ensure that differences in model performance were not driven simply by representation size. The same truncated SVD procedure was applied to the embeddings themselves, retaining their original number of dimensions (1024 for ProtBERT, 1280 for ESM2). This ensured that any effects of SVD transformation were mirrored in the embeddings, enabling a fairer comparison between OHE-SVD and the corresponding SVD versions of the pre-trained embeddings. Finally, to leverage complementary information captured at different levels, a combined embedding version was created for both ProtBERT and EMS2 by concatenating their corresponding global sequence embedding (sequence-level features) and the amino acid embedding (amino acid-level features), followed by applying SVD retaining the dimension number of each model embedding. This integration of distinct biological signals was hypothesized to enhance the model's learning capacity. The results from each classifier across all representation formats are presented in Table 1 for ProtBERT and Table 2 for ESM2.

**Table 1 – Test metrics for the different classifier models using ProtBERT embeddings**

| Representation | | Model | Accuracy | Precision | Recall | F1 Score |
|---|---|---|---|---|---|---|
| **Global sequence embeddings** | | **Random Forests** | 0.865 | 0.865 | 0.876 | 0.870 |
| | | **Logistic regression** | 0.854 | 0.849 | 0.873 | 0.861 |
| | | **MLP** | 0.871 | 0.872 | 0.889 | 0.876 |
| | **SVD** | **Random Forests** | 0.846 | 0.847 | 0.857 | 0.852 |
| | **SVD** | **Logistic regression** | 0.913 | 0.909 | 0.926 | 0.917 |
| | **SVD** | **MLP** | 0.933 | 0.934 | 0.937 | 0.936 |
| **Projected embeddings** | | **Random Forests** | 0.569 | 0.553 | 0.872 | 0.677 |
| | | **Logistic regression** | 0.518 | 0.518 | 1.000 | 0.682 |
| | | **MLP** | 0.513 | 0.449 | 0.867 | 0.591 |
| | **SVD** | **Random Forests** | 0.598 | 0.578 | 0.827 | 0.681 |
| | **SVD** | **Logistic regression** | 0.615 | 0.619 | 0.670 | 0.643 |
| | **SVD** | **MLP** | 0.610 | 0.610 | 0.687 | 0.646 |
| **Amino acids embeddings** | | **Random Forests** | 0.867 | 0.866 | 0.878 | 0.872 |
| | | **Logistic regression** | 0.851 | 0.843 | 0.876 | 0.859 |
| | | **MLP** | 0.897 | 0.896 | 0.912 | 0.901 |
| | **SVD** | **Random Forests** | 0.860 | 0.860 | 0.872 | 0.866 |
| | **SVD** | **Logistic regression** | 0.922 | 0.918 | 0.933 | 0.925 |
| | **SVD** | **MLP** | 0.938 | 0.938 | 0.942 | 0.940 |
| **Concatenated embeddings (amino acids + global)** | **SVD** | **Random Forests** | 0.855 | 0.854 | 0.868 | 0.861 |
| | **SVD** | **Logistic regression** | 0.922 | **0.918** | 0.933 | 0.926 |
| | **SVD** | **MLP** | 0.938 | 0.938 | 0.942 | 0.940 |
| **OHE** | **SVD** | **Logistic regression** | **0.930** | 0.917 | **0.950** | **0.933** |
| | **SVD** | **Random Forests** | **0.923** | **0.908** | **0.947** | **0.927** |
| | **SVD** | **MLP** | **0.946** | **0.947** | **0.948** | **0.947** |

Values represent the average performance across 30 independent data splits. Standard deviations were omitted for simplicity but are provided in Supplementary Table 2. SVD transformations in these experiments retained 1024 components to match the number of dimensions of ProtBERT embeddings. Bolded values denote the highest performance within each metric for each model, while underlined values indicate statistically significant differences in representation formats within each model, as determined by the Wilcoxon signed-rank test ($p < 0.05$). Note that recall values of 1.000, obtained in some cases when using the projected embedding, were disregarded as being the best since they result from poor learning performance in which all sequences were incorrectly assigned to the same class.

**Table 2 – Test metrics for the different classifier models using ESM2 embeddings**

| Representation | | Model | Accuracy | Precision | Recall | F1 Score |
|---|---|---|---|---|---|---|
| Global sequence embeddings | | Random Forests | 0.876 | 0.881 | 0.898 | 0.887 |
| | | Logistic regression | 0.802 | 0.805 | 0.828 | 0.815 |
| | | MLP | 0.831 | 0.837 | 0.874 | 0.849 |
| | SVD | Random Forests | 0.873 | 0.877 | 0.893 | 0.883 |
| | SVD | Logistic regression | 0.932 | 0.938 | 0.950 | 0.941 |
| | SVD | MLP | 0.949 | 0.950 | 0.955 | 0.952 |
| Amino acids embeddings | | Random Forests | 0.871 | 0.875 | 0.891 | 0.881 |
| | | Logistic regression | 0.806 | 0.813 | 0.842 | 0.824 |
| | | MLP | 0.888 | 0.894 | 0.911 | 0.899 |
| | SVD | Random Forests | 0.866 | 0.869 | 0.883 | 0.874 |
| | SVD | Logistic regression | 0.929 | 0.935 | 0.946 | 0.937 |
| | SVD | MLP | 0.946 | 0.947 | 0.952 | 0.949 |
| Concatenated embeddings (amino acids + global) | SVD | Random Forests | 0.868 | 0.872 | 0.888 | 0.878 |
| | SVD | Logistic regression | **0.933** | **0.939** | **0.950** | **0.942** |
| | SVD | MLP | 0.948 | 0.950 | 0.955 | 0.951 |
| OHE | SVD | Random Forests | **0.917** | **0.930** | **0.949** | **0.933** |
| | SVD | Logistic regression | 0.907 | 0.922 | 0.947 | 0.927 |
| | SVD | MLP | **0.950** | **0.952** | **0.957** | **0.953** |

Values represent the average performance across 30 independent data splits. Standard deviations were omitted for simplicity but are provided in Supplementary Table 3. SVD transformations in these experiments retained 1280 components to match the number of dimensions of ESM2 embeddings. Bolded values denote the highest performance within each metric for each model, while underlined values indicate statistically significant differences in representation formats within each model, as determined by the Wilcoxon signed-rank test ($p < 0.05$).

The first observation was the inferior performance when using projected embedding representation across all models, indicating that this embedding is less effective in protein modeling compared to its utility in NLP. A second observation was that the SVD transformation refined both ProtBERT and ESM2 embeddings, making them more effective for the task at hand, here, protein viability prediction, for both logistic regression and MLP, but not for random forests. Among pLM embeddings, amino acid-level representations yielded the best overall performance across models. Concatenating amino-acid and global sequence embeddings did not meaningfully improve results when using ProtBERT embeddings but led to gains with ESM2. This demonstrates the importance of local sequence features for protein classification and shows that incorporating global sequence information can provide additional predictive value depending on the pLM used. Unlike in NLP, where the corresponding embeddings are typically not used, the amino acid-based embeddings seem to provide the most valuable information in proteins. This aligns with the original ProtBERT and ESM2 studies, which recommend using amino acid-level embeddings for protein-level classification (or regression) tasks. However, the best overall performances were achieved with the OHE-SVD, although differences compared to the second-best representations were generally small, except in the case of random forests, where SVD-transformed ProtBERT embeddings did not lead to substantial improvements.

## 2.3. Mutation landscape analysis of functional groups

Training three different classifiers across multiple representation formats produced several model-representation pairs that offered a valuable opportunity to explore the biological features captured by each representation format. One dataset partition was randomly selected to investigate this, and the corresponding test set prediction accuracy was evaluated for each model-representation combination. Based on these predictions, sequences were grouped according to their classification outcomes to better understand which inputs were consistently easy or difficult to classify, and which were always well-classified or misclassified for the same representation. The groups were defined as follows: (i) sequences correctly predicted by all model–representation pairs (easy), (ii) sequences misclassified by all model–representation pairs (difficult), and sequences correctly classified or misclassified by all model–representation pairs using (iii) OHE, (iv) global sequence embeddings, (v) projected embeddings, or (vi) amino acid embeddings. Analyzing classification patterns across these groups can reveal the specific strengths and limitations of each representation in capturing biologically meaningful features.

The analysis began by examining the composition of each group in terms of viability and design strategy (Figure 4). All sequences labeled as *easy* turned out to be viable, while none was labeled as *difficult*. Yet, if using ProtBERT only, a few difficult sequences indeed emerged (Supplementary Figure 3), and these were all non-viable. Systematic misclassification with projected embeddings was observed only for non-viable sequences. In contrast, OHE, global sequence embeddings, and amino acid embeddings did not display any clear misclassification bias toward either class. Regarding design strategy, easy sequences were predominantly enriched in those generated by ML-based methods, whereas difficult sequences originated from both approaches. This observation aligns with the broader trend that consistent correct predictions were more frequent for ML-designed sequences.

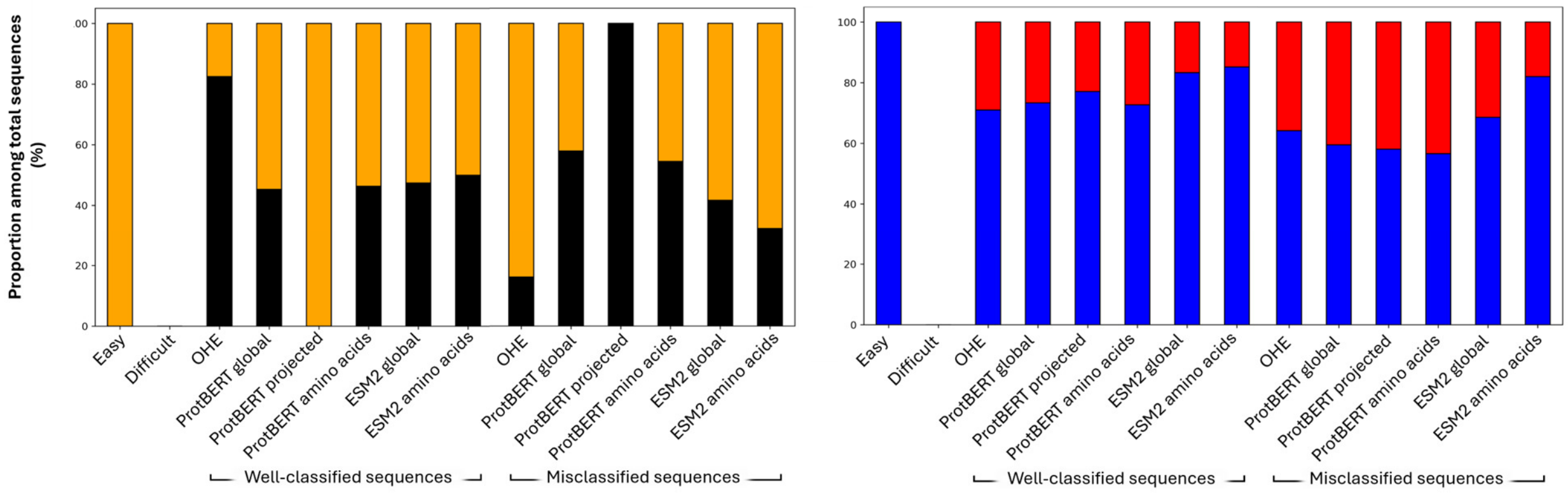

**Figure 4.** Composition of sequence groups. Groups of sequences sharing functional features (classification difficulty and correctness across all model-representation pairs featuring the specified representation format) are shown in terms of viability (left panel) and design strategy (right panel).

To further investigate the origin of these outcomes, the mutational landscape of each group, i.e., the frequency and type of mutations at each position within the targeted region, relative to the reference sequence, was analyzed. This analysis was conducted for the following groups (Figure 5): i) viable or non-viable sequences, ii) ML-designed or non-ML-designed, and iii) easy or difficult sequences. For difficult sequences, we considered those that emerged when using classifier–representation pairs based on ProtBERT embeddings only, since once pairs incorporating ESM2 embeddings were included, all sequences were correctly classified by at least one classifier–representation combination; in this case, correct classification was necessarily achieved by one or both of ESM2-based pairs. Results revealed that ML-designed and non-ML-designed sequences shared similar mutation profiles regarding the targeted positions and types of mutations at those positions, although ML-designed sequences exhibited a higher mutation load per position. This aligns with what is known about this dataset, since ML was used in the original study to guide extensive sequence diversification, thus leading to these sequences having more changes relative to the reference sequence compared to non-ML-designed sequences, as described by Bryant *et al.* (2021) [5]. Concerning viability, the analysis revealed a clear signature: viable sequences tend to avoid mutations within the region spanning amino acids 567–576. When mutations do occur in this region, they are predominantly substitutions. This region corresponds to a portion of the protein that is more buried within the capsid structure (data not shown) and is therefore likely critical for maintaining capsid integrity, providing a plausible structural explanation for the observed mutation avoidance. This mutational signature helps explain the differences between easy and difficult sequences. Easy sequences are all viable, and they do display the expected mutational landscape typical of viable variants. Difficult sequences, on the other hand, exhibit a mutational landscape similar to that of viable sequences but ultimately prove non-viable, which likely accounts for the classification challenge. This highlights the limitations of relying solely on positional encoding. While mutation type and location are considered, the specific amino acid substitutions and their resulting physicochemical properties, which critically affect protein function, are not taken into account. This *difficulty* disappears when ESM2-base representations are introduced suggesting that ESM2 embeddings encode functionally salient constraints, likely related to evolutionary and structural compatibility of amino acid substitutions, that are not sufficiently captured by ProtBERT. We also explored the mutational landscape across individual representation formats groups; however, no clear mutational signatures were identified that were specifically linked to any particular representation type (Supplementary Figure 4).

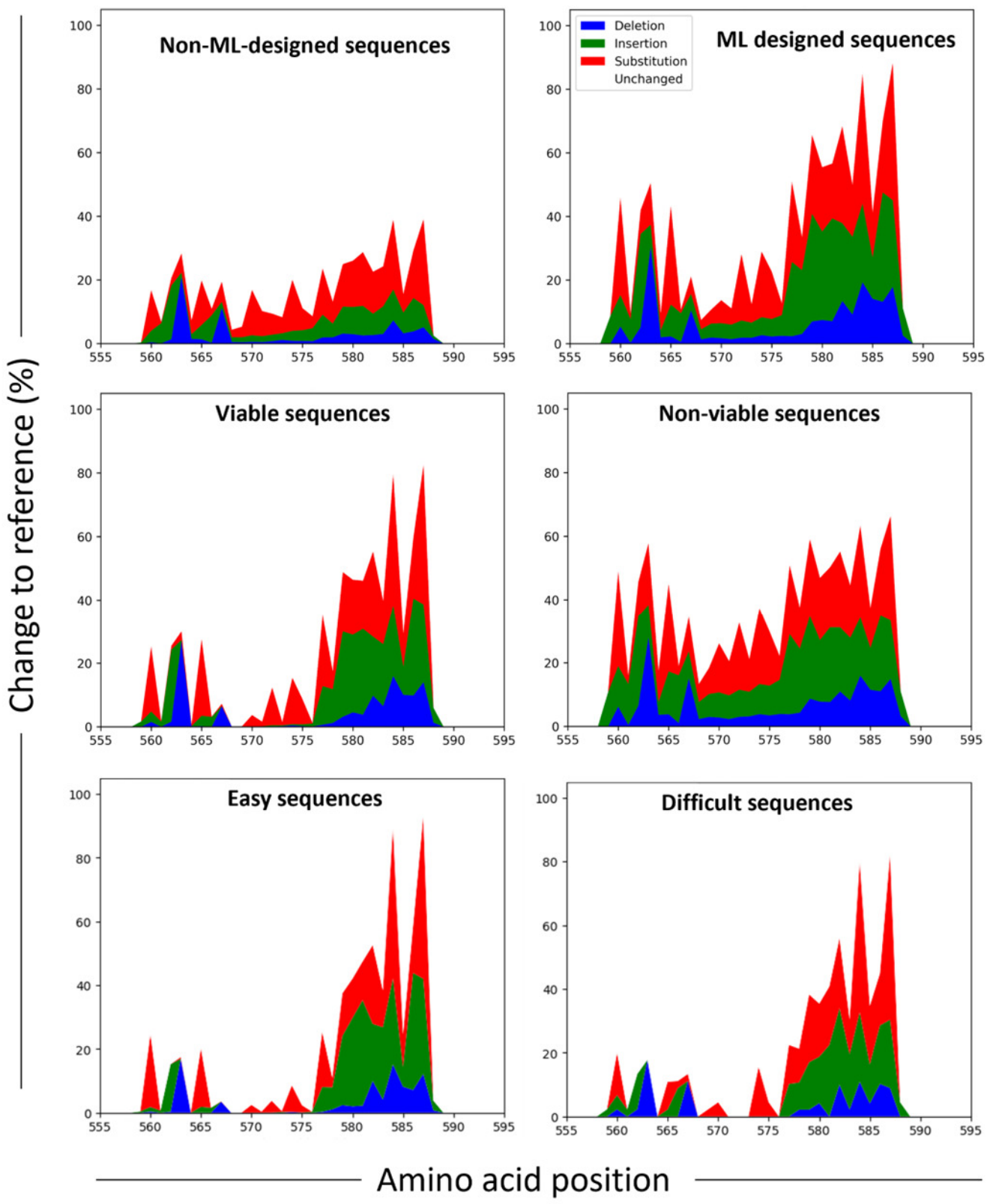

**Figure 5.** Mutational landscape analysis. Distribution of mutation types (deletions, insertions, and substitutions) across amino acid positions 561 to 588 in the targeted region, expressed as the percentage of change relative to the reference sequence. Data are presented for the following sequence groups: non-ML-designed, ML-designed, viable, and non-viable, easy and difficult sequences. For difficult sequences, we considered those that emerged when using classifier–representation pairs based on ProtBERT embeddings only.

### 2.4 Impact of mutation number and localization on different representation formats

The limitations of general-purpose pLM embeddings found in the AAV2 bioengineering study are not specific to this dataset. We demonstrate their broader relevance with an unrelated deep mutational scanning dataset of the SARS-CoV-2 spike protein [43] (Supplementary Figure 5). Embeddings generated with pLMs struggle to capture meaningful distinctions when mutations are either sparsely distributed across the entire sequence or densely concentrated within short, localized regions. This is likely because sparse mutations introduce subtle, long-range effects on structure or function that sequence-based embeddings fail to register. In contrast, clusters of mutations generate highly context-dependent local features that are frequently diluted or lost when representations rely on global pooling or averaged embeddings [44,45]. However, systematic studies that directly evaluate how the extent and localization of mutations limit the ability of embeddings to distinguish between protein variants, and at which thresholds these limitations become critical, remain scarce. Therefore, we designed controlled mutation schemes for our case study protein, introducing mutations either sparsely across the whole sequence or concentrated within localized regions. Representations of these sequences, based on each of the ProtBERT or ESM2 embeddings variants as well as OHE, were analyzed using t-SNE (Figure 6). In the dispersed mutation setting, differences became apparent after 50-100 mutations, with complete separation between clusters observed only at 500 mutations. This separation was consistently visible in ProtBERT and EMS2 embeddings but absent in OHE. The application of SVD transformations did not provide additional benefits for any representation type in this dispersed mutation setting, as the transformation primarily spread data points more evenly across the plots but did not improve grouping by mutation extent. In the localized mutation setting, grouping by mutation extent was also detectable with global sequence embeddings and amino acid embeddings; however, it was far less pronounced than in the dispersed mutation setting, and complete group separation was never achieved, except with OHE-SVD. This representation format performed remarkably well in separating sequences by targeted fragment size, with the SVD transformation being essential for this outcome, as OHE alone produced no separation. In both cases (dispersed mutations and concentrated regions) distinct group separation only became apparent at mutation levels rarely encountered in experimental datasets (more than 500 dispersed mutations) whereas for localized changes, even sequences with the 248-residue targeted fragment did not form a separate group, except when using ESM2 embeddings which provide slightly better clustering, although also not a complete separation. These results demonstrate the challenge of sequence representation in protein bioengineering, where achieving such extensive or heavily targeted mutations is impractical, prohibitively expensive, and usually unnecessary, as only a few changes are often sufficient to achieve the desired functional changes.

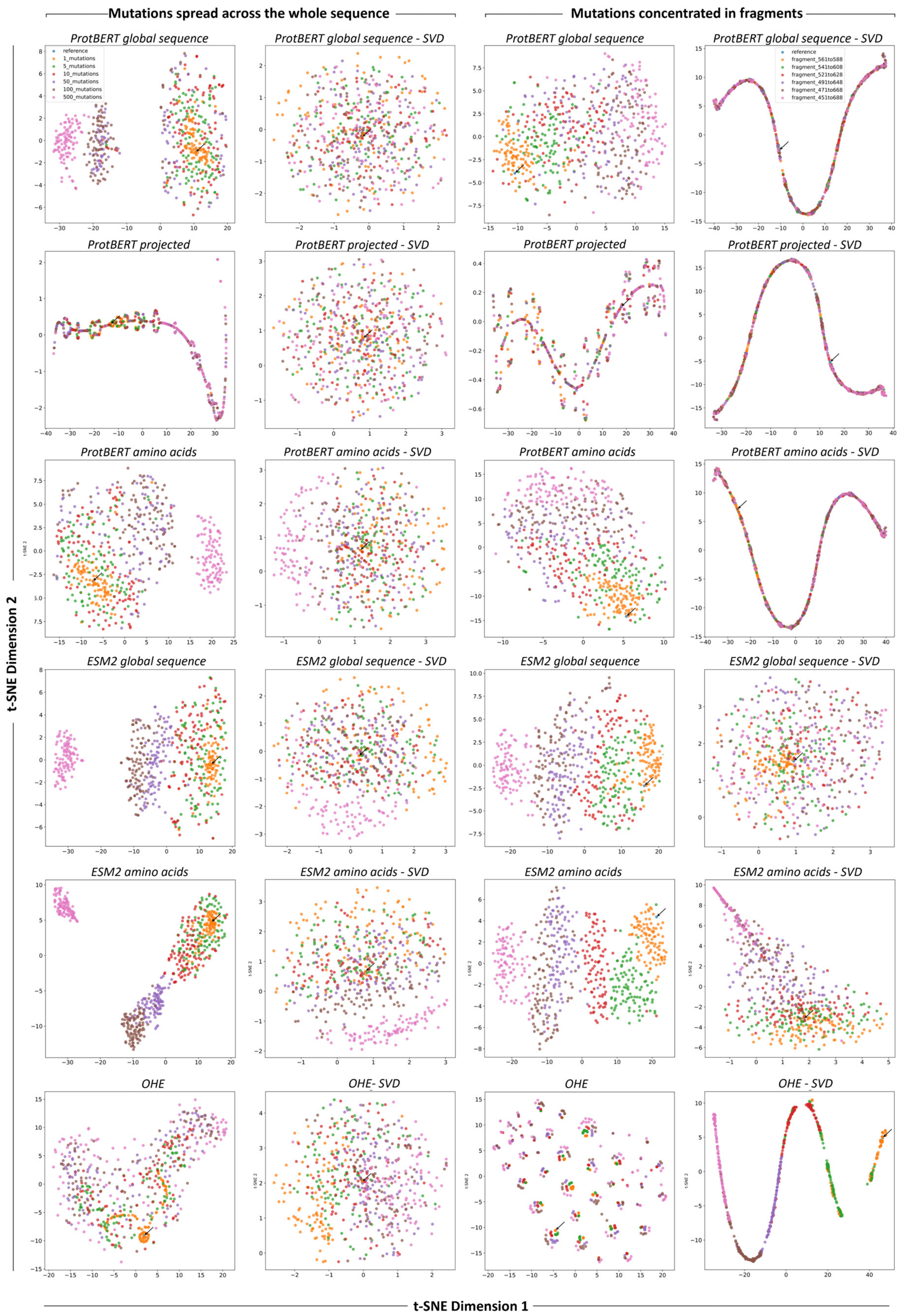
Mutations spread across the whole sequence
Mutations concentrated in fragments
ProtBERT global sequence
ProtBERT global sequence - SVD
ProtBERT projected
ProtBERT projected - SVD
ProtBERT amino acids
ProtBERT amino acids - SVD
ESM2 global sequence
ESM2 global sequence - SVD
ESM2 amino acids
ESM2 amino acids - SVD
OHE
OHE- SVD
OHE - SVD
reference
1_mutations
5_mutations
10_mutations
50_mutations
100_mutations
500_mutations
fragment_561to588
fragment_541to608
fragment_521to628
fragment_491to648
fragment_471to668
fragment_451to688
t-SNE Dimension 2
t-SNE Dimension 1

**Figure 6.** t-SNE plots of ProtBERT and ESM2 embedding variants and OHE representations featuring sequences with mutations distributed across the whole sequence (left panels) or concentrated in fragments targeting defined regions (right panels). For dispersed mutations, the indicated number of mutations was randomly introduced into the sequence, whereas for targeted fragments, every position within the specified regions was randomly mutated. For each representation, the corresponding SVD-transformation is also shown. The number of components retained in the SVD transformation was 1024 for ProtBERT embeddings and OHE, and 1280 for ESM2 embeddings. For simplicity, the 1280-component OHE-SVD representation is given in Supplementary Figure 6, as the conclusions do not change. In all panels, arrows indicate the reference (unmutated) sequence.

### 2.5 Task-specific fine-tuning

In pursuit of improved performances, we investigated whether supervised fine-tuning of ProtBERT and ESM2, using AAV2 capsid sequences labeled for viability, could enhance learning outcomes compared to their non-fine-tuned counterparts. Fine-tuning allows the embeddings to be adjusted for the specific prediction task, making small, localized differences more influential than the broader sequence-level patterns that might otherwise dominate in pre-training. To achieve this, we introduced a new architecture in which a linear neural network layer is added as the final component of the ProtBERT or ESM2 encoder, tailored for binary classification. This end-to-end setup eliminates the need for separate classification models, as the network produces the classification output directly. It also allows the loss to be backpropagated from the classification layer to the encoder, enabling the viability signal to directly guide the adaptation of representations for the specific task of viability classification. Fine-tuned embeddings improved learning performance across all variants examined, either surpassing the best overall results previously achieved using OHE for ProtBERT (Table 3) or achieving similar performances for ESM2 (Table 4). Moreover, fine-tuned embeddings demonstrated a clear separation of viable and non-viable sequences in the feature space, as evident in the t-SNE visualization (Figure 7), but not for other labels, such as sequence design strategy. This shows that fine-tuning enhances the discriminative power of the representations by aligning them more closely with the task-specific features, in this case, learning sequence viability. Here, the global sequence embedding format achieved the best performance among the fine-tuned variants for both ProtBERT and EMS2 embeddings, although in the later the differences were not statistically significant.

**Table 3 – Test metrics with fine-tuning models using ProtBERT embeddings**

| Representation | Accuracy | Precision | Recall | F1 |
|---|---|---|---|---|
| **Global sequence embeddings** | **0.955** | **0.958** | 0.955 | **0.957** |
| **Projected embeddings** | 0.952 | 0.948 | **0.959** | 0.954 |
| **Amino acids embeddings** | 0.947 | 0.941 | 0.958 | 0.949 |

Values represent the average performance across 10 independent data splits. Standard deviations were omitted for simplicity but are provided in Supplementary Table 4. Bolded values denote the highest performance, while underlined values indicate statistically significant differences as determined by the Wilcoxon signed-rank test ($p<0.05$).

**Table 4 – Test metrics with fine-tuning models using ESM2 embeddings**

| Representation | Accuracy | Precision | Recall | F1 |
|---|---|---|---|---|
| **Global sequence embeddings** | **0.951** | **0.952** | **0.955** | **0.953** |
| **Amino acids embeddings** | 0.949 | 0.951 | 0.952 | 0.951 |

Values represent the average performance across 10 independent data splits. Standard deviations were omitted for simplicity but are provided in Supplementary Table 5. Bolded values denote the highest performance, while underlined values indicate statistically significant differences as determined by the Wilcoxon signed-rank test ($p<0.05$).

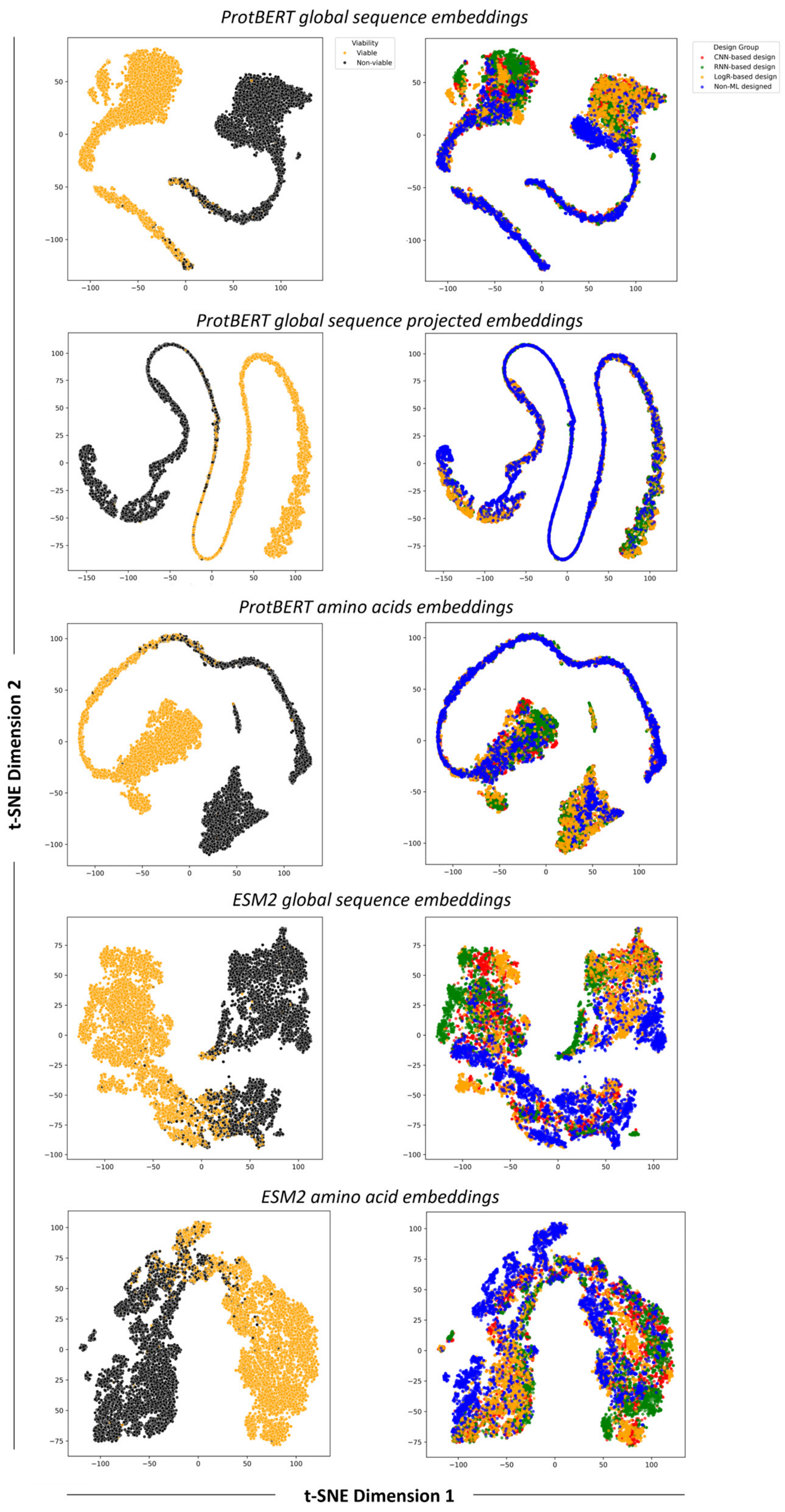

**Figure 7.** t-SNE plots of the different ProtBERT and ESM2 embedding variants after fine-tuning, colored by viability (top panels) or by design strategy (bottom panels).

The dynamics of fine-tuning–induced changes were further examined for each embedding variant (Figure 8). Because embedding dimensions are not directly comparable across model checkpoints, analyses were performed at the level of embedding subspaces. Differences between pre- and post-fine-tuned representations were quantified after aligning the embedding spaces and decomposed into orthogonal components. The analysis focused on the top components, which capture the dominant directions of change. A 95% cumulative variance threshold was chosen as it retains most of the signal while variance beyond this point increasingly reflects noise. For ProtBERT, one component (global and projected sequence embeddings) or two components (for amino acid embeddings) captured ≥95% of the change, whereas ESM2 embeddings required three (for amino acid embeddings) or seven components (for global sequence embeddings). This indicates that ESM2 distributes the task-specific adjustments across slightly more directions than ProtBERT, reflecting a more complex adaptation. Yet, in both cases, these components occupy only a small fraction of the full embedding space, demonstrating that fine-tuning affects a limited portion of embedding dimensions.

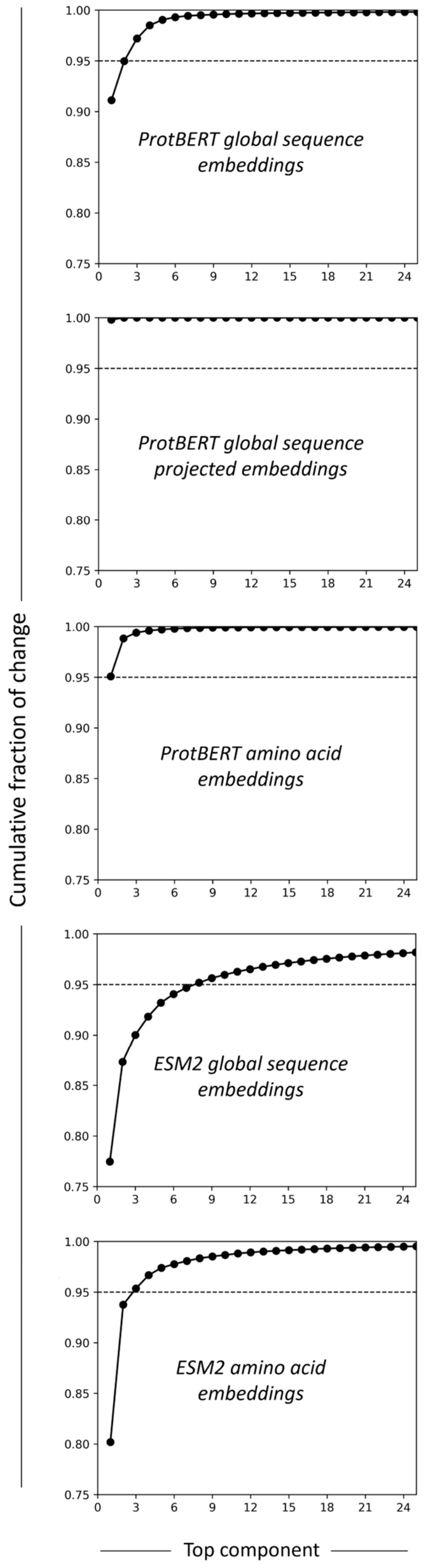

**Figure 8.** Cumulative variance explained by the top components of fine-tuning-induced changes in the embedding space for different embedding types. The dashed line indicates the 95% cumulative variance threshold.

## 3. Discussion

Embeddings generated with pLMs are currently a leading choice as ML-friendly formats to represent protein sequences due to their ability to capture rich, high-dimensional information about sequence context and short- to long-range interactions between amino acids [11]. However, pre-trained embeddings are inherently general-purpose representations that capture the broad grammar of protein language. On the other hand, tailoring protein functions in bioengineering is often achieved through small, localized changes in key regions of the sequence, since exhaustive mutation of entire sequences is neither feasible nor necessary to achieve functional changes. This challenges the effectiveness of pre-trained embeddings as representation formats, since such minor modifications are not sufficiently reflected in major embedding differences. To investigate this issue, we used the AAV2 capsid bioengineering case study, leveraging an extensive and high-quality dataset of sequence variants containing mutations confined to a small region of the protein, and evaluated the performance of pre-trained embeddings, specifically those of ProtBERT and ESM2, as protein representations in both unsupervised and supervised learning settings.

pLMs can generate multiple embedding variants, each potentially capturing different aspects of protein information, yet systematic evaluation of their comparative performance is lacking. Addressing this gap is important, especially as these models originate from the NLP domain. While analogies can be drawn between protein sequences and sentences, fundamental differences may limit the transferability of specific representation strategies [46]. In the unsupervised learning setting (Figures 2 and 3), the goal was to assess whether any representation format could naturally cluster sequences according to ground truth labels (viability and design strategy) and provide advantages in capturing these distinctions. Complete separation was not achieved with any of the representations, nor was it expected, particularly for viability, which is a complex biological outcome shaped by subtle, potentially non-linear interactions. Clustering was more evident for the design strategy, probably because machine-guided design introduces structured patterns that are more readily captured in the representations, enabling easier distinction in an unsupervised setting. Here, global sequence embeddings yielded the best functional clusterings while OHE primarily clustered sequences by length. This likely reflects that sequence length, although implicitly represented in embeddings, is more diffuse, whereas in OHE it is explicitly encoded and can dominate the representation (Supplementary Figure 1).

After exploring unsupervised learning approaches, the focus shifted to supervised learning, arguably the most directly applicable paradigm in machine-guided protein design and bioengineering. This is because many key objectives in this field, such as predicting the functional impact of mutations, designing novel protein variants with optimized properties, and classifying sequences according to specific phenotypic or biochemical traits, require modeling the sequence-to-function fitness landscape [47], which falls within the supervised learning paradigm. Here, one of the first notable findings was the consistently poor performance of ProtBERT projected embeddings across all tested models (Table 1). At first glance, this was surprising, considering the success of similar transformations in the NLP field, where dense layers followed by a tanh activation often enhance representations for tasks such as sentence classification. However, in the original BERT architecture, the pooler layer (from where the projected embedding is derived) was optimized for next sentence prediction objective, serving as a compact sequence-level representation [30], whereas ProtBERT was trained on masked language modeling [29]. Also, the tanh nonlinearity bounds each dimension of the embedding between −1 and 1, smoothing the representation by compressing high-magnitude activations. While this can improve stability and reduce noise for certain tasks, it may also discard variations in the global sequence embedding that are informative for sequence classification. Therefore, the projected embeddings derived from ProtBERT may not capture features relevant for classification tasks, and further smoothing can even obscure signals associated with small changes.

In the supervised learning setting (Table 1 and Table 2), embedding-based representations generally underperformed compared to OHE-SVD in most cases, although the differences were minor when considering the SDV-transformed formats, which is the fairest comparison, since only the SDV-transformed version of OHE was used here. This outcome is likely related to general-purpose pLM embeddings such as those of ProtBERT and ESM2 being trained to learn general representations of proteins rather than features directly tied to specific functional properties. While embeddings capture rich contextual and biological information that is useful for assessing overall sequence similarity, they do not inherently prioritize regions of the sequence that are most relevant to a specific task, such as viability prediction. In contrast, OHE provides representation that directly encodes position-specific sequence information without relying on learned global generalizations. This can be advantageous for tasks where only a small subset of positions drives functional outcomes, since the signal from these critical positions is preserved and accessible to the model. Interestingly, this advantage appears to derive from the SVD transformation combined with the focus on concentrated changes in the sequence, as OHE alone did not outperform pLM embeddings in the controlled mutation experiments (Figure 6).

An interesting finding of our study was that amino-acid embeddings averaged across the sequence generally outperformed global embeddings, despite the task involving only 28 mutated residues out of 735 (Tables 1 and 2). This likely reflects the fact that global embeddings emphasize overall sequence context and may downplay small, localized changes, whereas per-residue embeddings retain information at every position, and averaging preserves the contribution of the mutated residues, enabling better capture of subtle local effects. Thus, for tasks driven by localized sequence changes, averaged amino-acid embeddings can reveal more informative than global sequence embeddings. After fine-tuning, however, this trend was reversed, with global embeddings supporting better performances (Tables 3 and 4). A plausible explanation is that global embeddings, being directly supervised by the task loss, receive focused gradient updates that allow emphasizing the most informative positions. In contrast, averaging per-residue embeddings distributes the task-specific signal across the entire sequence, diluting the contribution of the critical mutated residues. Importantly, these two findings are not contradictory but reflect differences between frozen and fine-tuned representations. In the frozen setting, amino-acid-level embeddings retain localized mutation signals that may be attenuated in the global embedding. However, during fine-tuning, the global is directly optimized by the task loss and can learn to selectively amplify informative residues, whereas averaging per-residue embeddings distributes gradients across all positions, reducing sensitivity to localized changes.

Fine-tuning allows the model to assign higher relevance to features most critical for the downstream task, improving the overall predictive performance. These changes, although likely reflected in all embedding dimensions to some extent, seem affect more a small subset of the embedding subspace, likely because those dimensions carry the most task-relevant signals and require substantial refinement to enhance predictive performance (Figure 8). Connecting these changes to specific amino acids or functional regions can provide valuable biological insight for future work. Promising approaches toward that end include perturbation-based analysis, where individual residues are systematically masked or mutated in silico and the resulting changes in embedding space are measured to identify positions driving task-specific adaptation. Attention-based methods in transformer models offer a complementary strategy, by analyzing attention weights from amino acid embeddings before and after fine-tuning to identify residues that contribute most to the task-specific representation. Integrating these approaches could afford fine-tuning-induced changes to be mapped onto biologically meaningful sites, such as active, binding, or conserved regions, providing a direct link between embedding dynamics and protein function.

Another interesting finding was the performance improvement achieved by applying SVD to the embeddings (Tables 1 and 2). Typically, pLM embeddings are used in their original form, as directly outputted by the model, without dimensionality reduction or transformation. Here, the SVD transformation was applied solely as a control step to ensure a fair comparison with OHE, which was used in its SVD-transformed form. Remarkably, this processing step led to substantial improvements in the performance of models trained on the SVD-transformed embeddings compared to their untransformed counterparts except when using random forests. One possible explanation is that SVD reduces dimensionality while removing noise and redundancy, effectively increasing the signal-to-noise ratio. This benefits models sensitive to high-dimensional, correlated inputs, such as linear regression or neural networks, and aligns well with models that exploit linear separability or continuous gradients. In contrast, random forests, which rely on hierarchical binary splits rather than linear combinations of features, may derive less benefit and even perform slightly worse, which happened in some cases, as weakly informative components discarded by SVD could otherwise contribute to individual tree predictions. Similar improvements were observed for both ProtBERT and ESM2 embeddings, indicating that the effect is not pLM-specific but reflects a general property of SVD dimensionality reduction of emphasizing informative components and reducing noise.

This study provides new insights into using ProtBERT and ESM2 embeddings for ML-based tasks in protein design. The results show that, prior to fine-tuning, amino acid-level embeddings outperform sequence-level embeddings when used for whole-protein classification and functional prediction. The results also revealed that performance can be highly sensitive to how embeddings or position-encoding vectors are transformed, with effects depending on the model used. More importantly, our study demonstrates that pre-trained embeddings, although effective at capturing overall sequence context, are limited when functional outcomes depend on a few small or highly localized sequence changes, a situation commonly encountered in protein bioengineering. In this context, alternative pLMs, such as SaProt [48] and ProSST [49], may also be considered. However, mutation effect prediction models typically incorporate structural information or mutation-aware training objectives, which limits their applicability when high-quality structural data are unavailable. Therefore, fine-tuning general-purpose embeddings remains a practical and effective strategy requiring solely sequence data, enabling the model to adapt global representations toward task-relevant positions, thereby improving predictive performance in task-specific settings, as also demonstrated in recent studies [50–52]. As an alternative to fine-tuning, more sophisticated pooling strategies during embedding generation are also being explored to address the inherent limitations of the global nature of pre-trained embeddings and better capture small, localized sequence variations [45,53].

## 4. Methods

### 4.1. Dataset, data preprocessing, and mutation landscape analysis

The data used in this work is part of a dataset published by Bryant *et al.* (2021) [5], which reports a comprehensive study for machine-guided AAV2 capsid diversification. The set comprises 296,968 variants of the AAV2 capsid protein, both viable (153,691) and non-viable (143,278), all of which have been produced and experimentally validated in the laboratory. Each sequence in the dataset falls into one of two broad categories: non-ML-designed and ML-designed sequences (Supplementary Table 1). Non-ML-designed sequences were created during the first stage of the study by Bryant *et al.* (2021) [5] and were used to train ML models to generate new, more diverse sequences (ML-designed sequences). The dataset was downloaded from the original publication venue given in the 'Data Availability' section. This file contains details of the targeted fragment and the modifications introduced in each variant. For this project, the data in this file were processed to reconstruct the complete capsid protein sequences, incorporating the respective mutated fragments. This processing involved joining the fragments with the pre- and post-fragment sequences to reconstruct the entire protein and capitalizing all amino acid letters, as the original file represented amino acids resulting from insertions or substitutions in lowercase.

After pre-processing for whole-sequence reconstitution, the dataset was examined for duplicates, which were removed. To facilitate duplicate elimination, some subsets of non-ML-designed sequences were totally excluded, including ‘previous_chip_viable’, ‘previous_chip_nonviable’, and ‘singles’, along with 319 sequences from the ‘random_doubles’ subset. Additionally, the ‘wild_type’ subset was removed, since it contains the sequence that, being the reference, should not be used for training. The ‘stop’ subset was also excluded, since these sequences introduce significant complexity in pre-processing due to uncertainties in post-fragment reconstruction, while contributing only 57 records to the dataset. No duplicates were found in ML-designed sequences. In total, 3135 sequences were removed, resulting in a final dataset of 293835 sequences used here. Each sequence in the dataset was analyzed against the reference sequence (https://www.ncbi.nlm.nih.gov/protein/P03135.2) for substitutions, deletions, and insertions using the *pairwise2* method of Biopython. This method performs pairwise sequence alignment, allowing customization of penalties for matches, mismatches, and gaps. The default scoring parameters in Biopython were used (match: +1 point, mismatch: 0 points, gap: 0 points). While these can be adjusted if the alignment score is used for downstream analysis, they were irrelevant here, as the focus was solely on assessing the number and nature of mutations per sequence, rather than evaluating alignment scores. The number and nature of mutations per position (mutation landscape) were analyzed using custom-made functions.

### 4.2. Sequence representation formats

This work employed two types of encoding: OHE and pLM embeddings from ProtBERT and ESM2 from which two or three variants and a derivative aggregate version were used. For OHE, custom function was applied to each sequence which executes the following steps: (i) converts the sequences into lists of individual amino acid characters, (ii) pads the sequences with the token 'X' (a symbol not representing any amino acid) up to a maximum length, which in this case was the longest sequence found across the entire dataset, (iii) defines the order of amino acids for the binary vector (ACDEFGHIKLMNPQRSTVWYX), and (iv) iterates over each amino acid in each sequence to generate its corresponding binary vector, appending these vectors sequentially. To reduce feature complexity, improve computational efficiency, and standardize the feature dimensions between OHE and pLM representations, dimensionality reduction was applied to the OHE data using SVD, namely, truncated SVD, a simplified version of standard SVD that only computes and retains the top $k$ singular values and their corresponding singular vectors, rather than the full decomposition [54]. Herein, the *TruncatedSVD* class of scikit-learn was used with $k$=1024 or $k$=1280 , i.e., the top 1024 or 1280 singular values were retained, depending on the comparison embedding being from ProtBERT or ESM2, respectively. SVD-transformed versions of pLM embeddings were also generated for comparison. Standardization was applied to all SVD-transformed data using the *StandardScaler* class of scikit-learn.

For ProtBERT embeddings generation, the ProtBERT-BFD (big fat database, [55]) version was used. Sequences were pre-processed to align with the method's requirements as follows: (i) all characters were converted to uppercase, and non-standard or ambiguous amino acids were replaced with an 'X' (although such amino acids were not present in the datasets, this step was included to adhere to best practices), and (ii) a space was inserted between each letter. After the sequence pre-processing, embeddings were generated using the BertModel and BertTokenizer from the *transformers* library, by Hugging Face [56]. Three types of embeddings were directly extracted from the model output: global sequence embeddings, extracted with the instruction 'output.last_hidden_state[:, 0, :]', that is, the embedding corresponding to the [CLS] token, which captures the overall context of the input sequence; projected embeddings, extracted with the instruction 'output.pooler_output', that is, the [CLS] token embedding transformed using a nonlinear function (tanh), iii) amino acid embeddings, extracted with the instruction 'output.last_hidden_state[:, 1:-1, :].mean(dim=1)', that is, the global average of the individual amino acid embeddings, which captures the overall information of the protein sequence, excluding the special tokens [CLS] and [SEP].

For ESM2 embeddings, we used the esm2_t33_650M_UR50D model from the Fair-ESM 2.0 library. Sequences were pre-processed and tokenized using the model's batch_converter, which converts a list of sequence records into token tensors suitable for input to the model. The token tensors were then passed through the ESM2 model, and embeddings were extracted from the final-layer representations (out["representations"][model.num_layers]). Two types of embeddings were retained for downstream analysis: (i) global sequence embeddings, corresponding to the [CLS] token and capturing a summary of the entire sequence (representations[i, 0, :]); and (ii) amino-acid embeddings, where per-residue embeddings were first selected by masking out padding and special tokens ([CLS] and [EOS]) (representations[i, residue_indices, :]) and then averaged across the sequence (per_residue.mean(dim=0)), yielding a single vector representing the sequence based on amino-acid–level information.

### 4.3. Machine learning

Unless otherwise specified, all ML methods were implemented using the *scikit-learn* library [57]. The pre-processed dataset was shuffled to ensure a balanced representation of the different sequence designs and split into train-validation-test sets (70% / 10% / 20%) using viability-based stratification. For supervised learning, 30 independent splits were generated to train separate models and assess performance variability across different data partitions. For fine-tuning, 10 splits were used.

Unsupervised learning techniques, specifically HAC and t-SNE, were employed for exploratory data analysis and visualization using a sample of the training set. The *AgglomerativeClustering* scikit-learn class was used with the default parameters. The *TSNE* scikit-learn class was used with the following modifications: (i) the number of iterations was increased from 1000 to 10000 to ensure better convergence and capture more complex patterns in the data, and (ii) the perplexity was raised from 30 to 50 to account for the large dataset size, enabling the algorithm to consider more neighbors when modeling the local structure of the data.

Supervised learning was used to train binary classifiers to categorize sequences as viable or non-viable. Here, the following classifiers were implemented: i) logistic regression (*LogisticRegression* scikit-learn class), ii) random forests (*RandomForestClassifier* scikit-learn class), and iii) multilayer perceptron *MLPClassifier* scikit-learn class). For each classifier, 30 models were trained corresponding to 30 different data splits, with hyperparameters optimized via grid search. The parameters explored in the grid search were as follows: for logistic regression, the type and strength of regularization; for random forests, the number of trees and the maximum tree depth; and for the multilayer perceptron, the size of the hidden layer and the learning rate. Model performance was evaluated using standard classification metrics, including accuracy, precision, recall, and F1 score.

### 4.4 Model-representation pairs analysis

The performance of each model–representation combination directly was analyzed to gain further insights into how different representation formats influence predictive outcomes. For each model-representation pair, prediction accuracy on the test set was evaluated, and sequences were then grouped based on consistency in prediction outcomes and analyzed for viability, design strategy, and mutation landscape. The groups were defined as follows: (i) sequences correctly predicted by all pairs (easy), (ii) sequences misclassified by all pairs (difficult), and sequences correctly or incorrectly predicted by all pairs using (iii) OHE, (iv) global sequence embeddings, (v) projected embeddings, or (vi) amino acid embeddings. Only the simpler (non-concatenated) embedding variants were included for this analysis, with SVD-transformed embeddings considered alongside their corresponding original (non-SVD) versions.

### 4.5 Controlled mutations

To assess the impact of mutation number and localization across different sequence representation formats, a controlled mutation scheme was applied to the AAV2 reference sequence. Mutations were randomly introduced either sparsely throughout the entire sequence or concentrated within specific regions. In the sparse setting, 100 variants were generated with 1, 5, 10, 50, 100, or 500 mutations distributed throughout the sequence. In the localized setting, mutations were confined to a 28-residue fragment (positions 561–588) that was progressively expanded in 40-residue increments (20 upstream and 20 downstream) up to a 248-residue region, with all positions in each fragment mutated. In both settings, mutations were randomly selected as insertions, deletions, or substitutions.

### 4.6 Fine-tuning

The pre-trained ProtBERT or ESM2 models were fine-tuned using a supervised learning approach with viability labels for task-specific adaptation implemented in Pytorch version 2.7.1. Input sequences were processed in batches of size 10. Binary cross-entropy was used as the loss function, and model optimization was guided by a StepLR learning rate scheduler, which reduced the learning rate by a factor of 0.05 after each epoch to promote stable convergence. Training was capped at five epochs, with early stopping implemented to prevent overfitting (training was halted if validation loss did not improve over three consecutive epochs). To examine how fine-tuning affected the representations, the embedding subspace most influenced by fine-tuning was evaluated. Because similarity metrics are invariant to rotations and other transformations of the embedding space, comparisons focused on the overall geometry rather than individual embedding dimensions. Embeddings were centered by subtracting, for each dimension, the mean value across all sequences. Post-fine-tuning embeddings were then aligned to the pre-trained embeddings using the orthogonal Procrustes method, which finds the rotation minimizing the overall difference between the two embedding matrices[58,59]. The difference between aligned post-fine-tuning and pre-trained embeddings was computed for each sequence, forming a change matrix capturing fine-tuning-induced modifications. This matrix was decomposed using SVD to identify the principal directions along which embeddings were modified. Singular values indicate the magnitude of change along each direction, and the fraction of total change captured by each direction was computed by squaring the singular values and normalizing by the sum of squared singular values. Cumulative variance explained plots were used to assess the number of components needed to explain fine-tuning-induced changes.

## Funding

This work was supported by FCT - Fundação para a Ciência e Tecnologia, I.P. under the LASIGE Research Unit, ref. UID/00408/2025, DOI https://doi.org/10.54499/UID/00408/2025, and partially supported by project 41, HfPT: Health from Portugal, funded by the Portuguese Plano de Recuperação e Resiliência. It was also partially supported by the CancerScan project which received funding from the European Union's Horizon Europe Research and Innovation Action (EIC Pathfinder Open) under grant agreement No. 101186829. Views and opinions expressed are however those of the author(s) only and do not necessarily reflect those of the European Union or the European Innovation Council and SMEs Executive Agency. Neither the European Union nor the granting authority can be held responsible for them. Pedro Cotovio, Laura Balbi, and Lucas Ferraz acknowledge Fundação para a Ciência e a Tecnologia for the PhD grants, respectively, 2022.10557.BD, 2024.01208.BD and 2025.04034.BD.

## Author contributions

AFR, JLF, and CP conceived the research idea and designed the study. AFR, JLF and LB implemented the code, conducted the experiments, performed data collection, visualization, and analysis. PC contributed with critical insights and domain expertise throughout the study. AFR and JLF prepared the original draft of the manuscript, and all authors contributed to revising the final version. CP provided overall supervision, project leadership, and secured funding to support the research.

## Data availability

The dataset used in this study was generated by Bryant *et al.* (2021) [5] and is available at https://www.nature.com/articles/s41587-020-00793-4. Scripts and code used to generate the results in this study are publicly available at https://github.com/liseda-lab/embCOMP.

## Competing interests

The authors declare that they have no competing financial interests or personal relationships that could have influenced the work presented in this paper.